# Sending femtosecond pulses in circles: highly non-paraxial accelerating beams

F. Courvoisier,* A. Mathis, L. Froehly, R. Giust, L. Furfaro,
P. A. Lacourt, M. Jacquot, J. M. Dudley

*Département d'Optique P.M. Duffieux, Institut FEMTO-ST,
UMR 6174 CNRS Université de Franche-Comté, 25030 Besancon Cedex, France*
*Corresponding author: francois.courvoisier@femto-st.fr*

We use caustic beam shaping on 100 fs pulses to experimentally generate non-paraxial accelerating beams along a 60 degree circular arc, moving laterally by 14 µm over a 28 µm propagation length. This is the highest degree of transverse acceleration reported to our knowledge. Using diffraction integral theory and numerical beam propagation simulations, we show that circular acceleration trajectories represent a unique class of non-paraxial diffraction-free beam profile which also preserves the femtosecond temporal structure in the vicinity of the caustic.

Transversally accelerating beams such as Airy beams exhibit a curved trajectory of their point of maximum intensity. Since the intensity maximum is strongly localized over an extended propagation distance along the trajectory, they constitute a novel class of "non-diffracting" beam and have attracted tremendous interest for applications in both linear and nonlinear optics [1-5]. The studies of such accelerating beams, however, has to date been confined only to the paraxial regime, for relatively small deviation angles typically less than 10 degrees.

In this letter, we apply a recently-developed caustic-based approach to accelerating beam synthesis [6,7] to generate accelerating beams in the non-paraxial regime. We consider the particular case of a circular trajectory, and accelerate 100 fs pulses at 800 nm along a 60 degree circular arc, moving a primary intensity lobe of 1 µm size laterally by 14 µm over a 28 µm propagation distance. To our knowledge, this is the highest degree of transverse acceleration reported to date. We also analyze the spatio-temporal beam properties using diffraction integral theory, and show how the size of the primary lobe is directly linked to the local radius of curvature of the trajectory. Moreover, we show that in the non-paraxial regime, it is only the circular propagation trajectory that can be considered as genuinely diffraction-free. We also use numerical beam propagation to show that the beam acceleration does not modify the temporal pulse duration in the vicinity of the primary lobe.

Our experiments applied caustic beam shaping to the Gaussian beam profile from a Ti:Sapphire laser operating at 800 nm and generating 100 fs pulses [7]. The collimated Gaussian beam (with near-uniform spatial phase) was incident upon an optically-addressed SLM (Hamamatsu PAL-SLM) to which an appropriate caustic-phase function was written. In contrast to Fourier-based approaches to generate accelerating beams, our approach applies the phase function directly to the incident beam.

The starting point with this approach is the desired acceleration trajectory $c(z)$ along the propagation direction $z$. With $c(z)$ defined, the corresponding SLM phase mask $\Phi(y)$ to apply in the transverse beam plane $y$ is determined from: $d\Phi/dy = kc'/(1+c'^2)^{1/2}$ where $k$ is the wave vector and the derivative $c' = dc/dz$. The beam characteristics in $z$ and $y$ are related using the Legendre transform of the caustic [7]. The beam was imaged and demagnified by a factor of 278 using a lens and a microscope objective (numerical aperture 0.8) in a 4f configuration to allow us to obtain caustic trajectories of micron dimensions (suitable for applications in micromachining and nonlinear optics). The plane at which the SLM is imaged defines the point from where the micron-scale accelerating beam is generated, and all results below show characterization of the demagnified beam after the SLM image plane. Note also that the beam polarization was perpendicular to the trajectory plane.

Figure 1 shows results obtained for a target non-paraxial trajectory corresponding to a circle of radius 35 µm. The results in the top subfigure of Fig. 1(a) compare the target caustic (dashed white line) with the results of the numerical propagation of a Gaussian beam to which an ideal caustic phase profile is applied. These numerical results confirm that we can indeed achieve non-paraxial acceleration with our setup. The results in the bottom subfigure of Fig. 1(a) provide experimental confirmation. We see very good agreement between experiment and the numerical results, both in terms of the transverse localization of the beam intensity along the target caustic trajectory, as well as the variation of the local intensity with propagation along the caustic (we discuss this aspect in more detail below). The residual scattering and fringes seen inside the caustic in experiment arise from unfiltered zero-order diffraction, but we see that this does not significantly modify the field localization along the caustic itself.

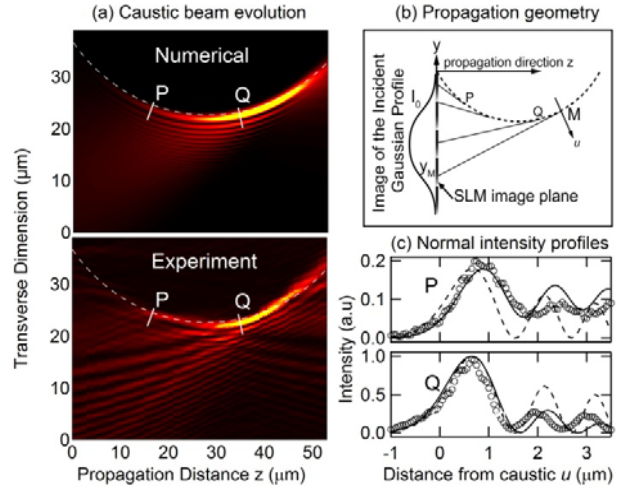

Fig. 1 (a) Numerical (top) and experimental (bottom) intensity distributions of a circular accelerating beam. The corresponding target caustic is shown as the white dashed line. (b) Propagation geometry of the caustic beam. (c) Comparison of the numerical (line) and experimental (circles) intensity profile normal to the caustic at points P and Q as indicated. In (c) we also show the analytic result from Eq. 1 (dashed line).

The experimental acceleration profile of a 35 µm radius circle is maintained over an arc exceeding 60°. The effective trajectory moves laterally by 14 µm over a propagation distance of only 28 µm, which is the highest degree of acceleration experimentally published to our knowledge. Note that although we show results here for a circle, other cases of non-paraxial caustics (e.g highly-non paraxial polynomials) have also been generated using our setup, and display similar characteristics.

To interpret these results further, Fig. 1(b) shows details of the caustic geometry beyond the SLM image plane, defining a normal direction to the caustic ($u$) along which we can extract and plot the local intensity profile at any point M. Figure 1(c) shows different intensity profiles along the caustic at two points P and Q, comparing numerical results (line) and experimental measurements (circles), with good agreement.

The different values of maximum intensity at P and Q and the intensity variation along the caustic arise from the use of finite numerical aperture Gaussian beam illumination. Figure 1(b) shows how this can be readily understood in terms of simple geometrical arguments. Specifically, the tangent rays from the SLM image plane corresponding to points P (low intensity) and Q (high intensity) arise respectively from points in the wings and near the center of the incident Gaussian beam. It is thus readily seen that it is the intensity difference between the two points on the initial beam that feeds in directly to the intensity profile along the accelerating profile.

The above qualitative description can be expressed quantitatively using analysis based on the non-paraxial Sommerfeld diffraction integral. Using this approach, we derive the intensity profile at any point in the vicinity of the caustic as follows:

$$I_M(u) = \left(\frac{2c''}{k}\right)^{2/3} \frac{kz}{2\pi\sqrt{1+c'^2}} I_0(y_M) \text{Ai}^2\left[-\left(\frac{2k^2}{R(z)}\right)^{1/3} u\right] \quad (1)$$

Here, $M = M(z, c(z))$ is the point at which the intensity $I_M(u)$ is calculated, and the normal coordinate $u$ is as defined in Fig. 1(b). Ai is the Airy function, $I_0(y_M)$ is the intensity of the incident beam corresponding to the ray from the SLM image plane tangent to $M$, and $R(z) = (1+c'^2)^{3/2}/c''$ is a local radius of curvature geometrically defined for every point on the trajectory. Primes indicate differentiation with respect to $z$.

The non-paraxial intensity given by Eq. (1) is compared to the experimental results (no free parameters) at the points P and Q in Fig. 1(c). The agreement at Q is very good in the vicinity of the primary lobe, but we see deviations for the subsidiary lobes and also for the point P. In fact, this is expected since Eq. (1) is derived assuming a slowly varying incident beam profile and the caustic at P and the side lobes are generated from points on the beam away from the slowly varying intensity maximum. Better agreement at all points on the caustic would be obtained with a larger input beam and SLM. We also note that the result expressed in terms of an Airy function is obtained using a third-order phase expansion which is valid for caustic trajectories with non-vanishing second derivatives $c''$. For such cases, our analysis extends into the non-paraxial regime the universal association of the Airy function with arbitrary acceleration profiles proposed in Ref. 6. Our analysis can be readily generalized to other classes of trajectory using higher order phase expansions (using e.g. Pearcy integrals), but the results are qualitatively similar to the Airy function in yielding beam properties in terms of a primary central lobe with lower amplitude oscillations [8,9].

We also use Eq. (1) to consider more generally the nature of "non-diffracting" beams in the non-paraxial regime. To this end, we derive the transverse size (FWHM) of the primary intensity lobe in the direction normal to the caustic as: $w(z) = \Delta u[R(z)/2k^2]^{1/3}$ with $\Delta u = 1.630$ the half maximum of the intensity $|Ai|^2$. From this result, we see that the condition for a "non-diffracting" trajectory [i.e. corresponding to a z-invariant $w(z)$] is only achieved when the local radius of curvature $R(z)$ is constant. This corresponds to requiring that the chosen trajectory $c(z)$ is in fact a circle, and a consequence of this is that all trajectories that are not circular, including the well-known parabolic beam trajectory used to initially generate accelerating beams, are in fact not ideally diffraction-free outside the paraxial regime.

This non-diffracting nature of the circular caustic is confirmed in Fig. 2(a) which plots the evolution with propagation distance of the primary lobe size $w(z)$, comparing experiment (circles) with the analytic prediction for $w(z)$ (dashed line), as well as results extracted from numerical beam propagation (solid line). The circular caustic maintains constant width ~1 μm over a significant distance of ~ 20 μm. For comparison, a Gaussian beam of the same size (as beam waist) would have a Rayleigh range of only 3.4 μm.

The shaded region in the figure is used to show the onset of the regime where numerical beam propagation indicates that the effects of finite numerical aperture of the incident beam induces deviation from diffraction-free behavior. Additional results are shown in Fig. 2(b) where the result in Eq. (1) for the variation of the intensity maximum along the caustic $I_M(u = 0)$ is compared with experimental and numerical results. We see good agreement in the diffraction-free regime, further providing confirmation of our analysis and interpretation.

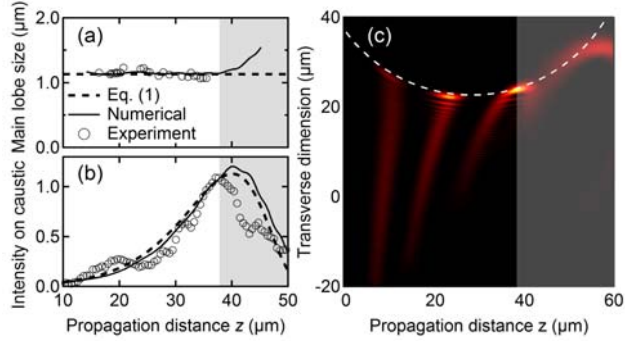

Fig. 2 (a) Evolution of the primary lobe size with propagation. The dashed line corresponds to the analytical model of Eq. (1), the solid line to the numerical propagation and the circles to experimental results. (b) Evolution of the intensity along the caustic with the propagation distance. (c) Snapshots of a 15 fs pulse at various points along its propagation. The corresponding propagation times are 40, 80, 140 and 230 fs.

The high acceleration in the non-paraxial regime raises a natural question of the effect on the spatio-temporal structure of the beam in the femtosecond regime [10]. We have numerically investigated this using the non-paraxial plane wave spectrum model of propagation, neglecting SLM material dispersion (valid for pulses longer than ~10 fs). At four different points on the caustic, Fig. 2(c) shows snapshots of the structure of a 15 fs pulse at 800 nm as it propagates. Figure 3 shows the spatio-temporal intensity plotted in a reference frame of co-moving time and spatial distance from the caustic parallel to the $y$ axis. This view clearly shows how the propagation along the caustic is associated with locally-varying pulse front tilt and spatio-temporal structure, but significantly, in the non-diffracting regime at $z = 15$ μm, 25 μm and 35 μm, the temporal structure of the primary lobe in the vicinity of the caustic intensity maximum is invariant. This is seen in the figure where we plot the profile on the target caustic (at $u = 0$) as the black curve. On the other hand, outside the non-diffracting regime for $z = 55$ μm, we see significant pulse temporal broadening. The pulse widths at these points are given in the figure.

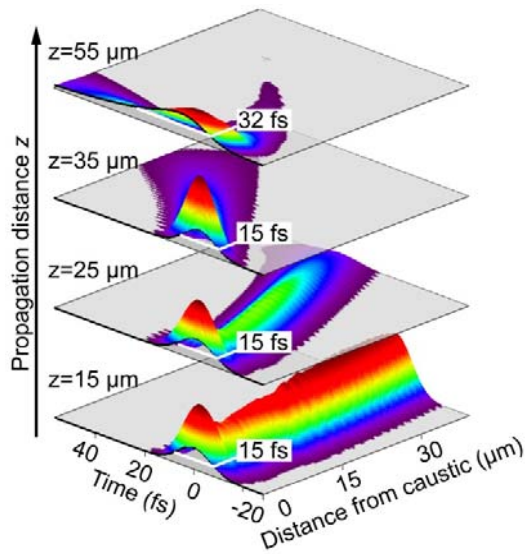

Fig. 3. Intensity distribution of a 15 fs pulse at propagation distances as shown as a function of co-moving time and distance from the caustic.

The major result of this paper has been to experimentally generate non-paraxial non-diffracting beams along a circular trajectory. Our results have been interpreted with numerical and analytical studies, with the latter providing insight into the nature of diffraction-free propagation in the non-paraxial regime. Within the diffraction free regime, the temporal structure at the caustic maximum is preserved, and we expect our results to impact on uses of highly accelerating beams in nonlinear optics.

We thank the Université de Franche Comté, the Région Franche-Comté and the Agence Nationale de la Recherche contract 2011-BS04-010-01 NANOFLAM.